# Ultrafast Processes in Isomeric Pyrene-*N*-methylacetamides: Formation of Hydrogen Bond Induced Static Excimers with Varied Coupling Strength

*Krishnayan Basuroy,*[*,§] *Jose de J. Velazquez-Garcia,*[§] *Darina Storozhuk,*[§] *David J. Gosztola,*[¥] *Sreevidya Thekku Veedu,*[§] *and Simone Techert*[*,§,£]

[§]Structural Dynamics in Chemical Systems, Photon Science Division, Deutsches Elektronen-Synchrotron DESY, Notkestraße 85, Hamburg, 22607, Germany

[¥]Center for Nanoscale Materials, Argonne National Laboratory, Lemont, Illinois 60439, United States

[£] Institute of X-ray Physics, University of Göttingen, Friedrich-Hund-Platz 1, 37077 Göttingen, Germany

AUTHOR INFORMATION

**Corresponding Author**

*E-mail: krishnayan.basuroy@desy.de (K.B.).

*E-mail: simone.techert@desy.de (S.T.).

**ABSTRACT.** Pyrene based molecules are inclined to form *excimers* through self-association upon photoexcitation. In this work, the pyrene core is functionalized with *N*-methylacetamide group at the position 1 or 2, to develop pyren-1-methylacetamide (PyMA1) and pyren-2-methylacetamide (PyMA2), respectively. Upon photoexcitation, PyMA1 and PyMA2, at 1.0mM, in toluene, have formed predominantly *static* excimers. The steady state spectroscopic studies have showed that the excitonic coupling of PyMA1 dimers are much stronger in solution than its isomeric counterpart, PyMA2. The transient absorption (TA) measurements over *fs-ps* regime (*fs*-TA) showed that the formation of static excimers with the strongly-coupled pre-associated dimers, in PyMA1, happens in ~560*fs*, whereas, the excimers for the weakly-coupled pre-associated dimers in PyMA2 have formed in much slower time scale (~65*ps*). The introduction of methylacetamide group at the position 1 or 2 on pyrene ring, was believed to have allowed forming hydrogen bonded excimers with different degrees of excitonic coupling.

**TOC GRAPHICS**

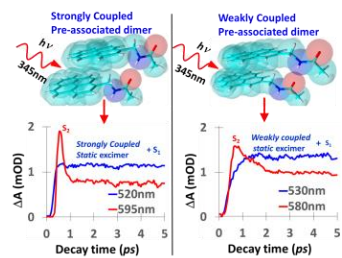

Pyrene based fluorophores have attracted appreciable amount of interest, owing to their long lived singlet excited state, high fluorescence quantum yields, good thermal stability and high charge carrier mobility. These features make pyrene based systems suitable for designing a wide range of organic electronic devices, such as organic light-emitting diodes (OLEDs), organic photovoltaics (OPVs), organic field-effect transistors (OFETs) and organic lasers.[1] However, a major shortcoming of using pyrene based luminescent materials is the extreme sensitivity of their fluorescent property towards fast non-radiative quenching of the monomeric fluorescence at close aggregation, a phenomenon, known as aggregation caused quenching (ACQ).[2] In pyrene based systems, very often, the ACQ of monomeric fluorescence is accompanied by the formation of aggregation induced excited state dimers, known as excimers.[3-5] Excimer formation drastically quenches the monomeric emission in the aggregation state and can function as trap states that restrict the targeted processes such as energy transfer and exciton diffusion, making it undesirable for energy conversion or light harvesting applications.[6-9] On the other hand, recent evidences have shown that excimer formation have facilitated the singlet fission process.[10-12] The excimer-induced enhanced emission (EIEE) in pyrene based systems also show that the excimers possess higher photoluminescence efficiency compared to their monomeric counterparts.[13,14] White light emitting (WLE) materials are also designed by exploiting the easy tunability of the well separated monomer and excimer emission bands across the entire visible wavelength range.[15,16] Pyrene based excimers are also utilized as fluorescent probes to detect the presence of transition metal complexes and examine protein-protein interactions.[17-19]

Depending on how they form, excimers can be divided into two classes, *static* and *dynamic*.[20] The static excimers are formed due to the photoexcitation of ground state dimers that are already pre-associated through π···π stacking or van der Waals interactions. The dynamic excimers are formed as a result of interaction between the molecules in lowest lying singlet excited state with identical molecules in the ground state. Recently, a $C_3$-symmetric molecular probe (PYTG) consisting pyrene and triaminoguanidinium chloride has been utilized to understand the static and dynamic excimer formation in a single molecule platform, where the dynamic excimer was formed by changing the concentration and the static excimer was formed in the presence of specific ions.[21]

In this letter, we have investigated two mono-substituted, isomeric pyrene derivatives, pyren-1-methylacetamide (PyMA1) and pyren-2-methylacetamide (PyMA2), where pyrene is functionalized by *N*-methylacetamide group at 1- and 2- positions, respectively (Figure 1a,b). As a conscious effort to utilize the intermolecular hydrogen bond interactions for keeping the relative orientations of neighboring pyrene moieties in a certain way during excimer formation,[22] we have employed methylacetamide to attach on the pyrene ring which contains both hydrogen bond donor (NH) and acceptor (CO) groups. The dynamics of pyrene excimers, either in gases, solutions or solids, is an ultrafast process, typically occurs in the picosecond or sub-picosecond time scale and makes it technically challenging to study them in real time situation, even while equipped with ultrafast laser sources and detectors with fast readouts.[23-25] We have employed ultrafast optical transient absorption spectroscopy technique in femtosecond-picosecond (*fs-ps*) and nanosecond-microsecond (*ns-μs*) regimes to examine the ultrafast dynamics of transient species followed by the laser excitation.

Initially, a comparative analysis of the UV-Vis absorption and fluorescence emission spectra collected for PyMA1 and PyMA2, at 0.1mM and 1.0mM concentrations was performed. At 0.1mM, both PyMA1 and PyMA2 have exhibited monomeric characteristics, with no sign of close aggregation in toluene (Figure 1c,d). However, for PyMA1 the UV-Vis spectrum at 1.0mM show strong signs of ground state dimerization with a drastic change in the relative vibronic peak intensities in the higher wavelength regime (Figure 1c). The absorption spectrum of PyMA2 at 1.0mM shows a small change in the relative intensities between the vibronic bands at higher wavelengths (Figure 1d). The fluorescence spectroscopy with 345nm excitation reveals the excimer formation for both the compounds at 1.0mM in toluene, with the appearance of bathochromically shifted, broad and structure-less band in the emission spectra (Figure 1c,d). Close examinations of the emission spectra showed that the excimer bands corresponding to PyMA1, centered at ~480nm, are slightly bathochromically shifted from the excimer band observed for PyMA2 centered at ~475nm. The relative intensity of excimer bands compared to monomer bands showed excimer formation is much more pronounced in PyMA1 than PyMA2.

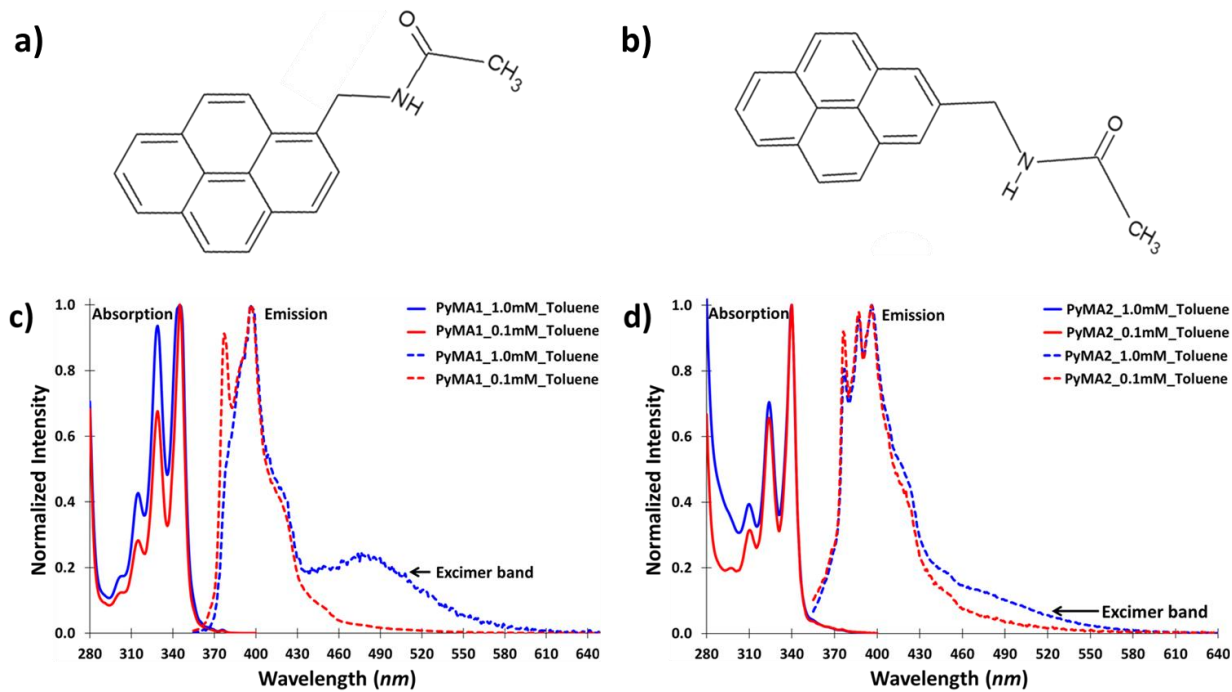

**Figure 1.** Chemical structure of a) PyMA1 and b) PyMA2. Normalized absorption (solid lines) and emission spectra (dashed lines) of c) PyMA1 and d) PyMA2 at 0.1mM (red) and 1.0mM (blue). $\lambda_{exc}$ = 345nm. Solvent = Toluene.

The fluorescent excitation spectra of PyMA1 also shows a significant bathochromic shift of the low energy excitation bands at 1.0mM compared to the bands at 0.1mM (Appendix Figure S6a). But for PyMA2, there is no shifting of the excitation bands from 0.1mM to 1.0mM. Although, the relative intensity of the different vibronic bands has suffered a small change, along with very little broadening (Appendix Figure S6b). The absorption, excitation and emission spectra collected at 1.0mM for PyMA1, indicates a strongly-coupled dimerization, both in ground and excited states, whereas for PyMA2, the coupling between the dimers are relatively weak.[26] The scheme 1a depicts the mechanism for the formation of static and dynamic excimers, respectively. In the present study both the systems are following the mechanism of static excimers. A hypothetical schematic, exhibiting the potential energy surfaces corresponding to the both kinds of static excimers formed in PyMA1 and PyMA2 are also illustrated with the help of steady state spectroscopy results (Scheme 1b, c).

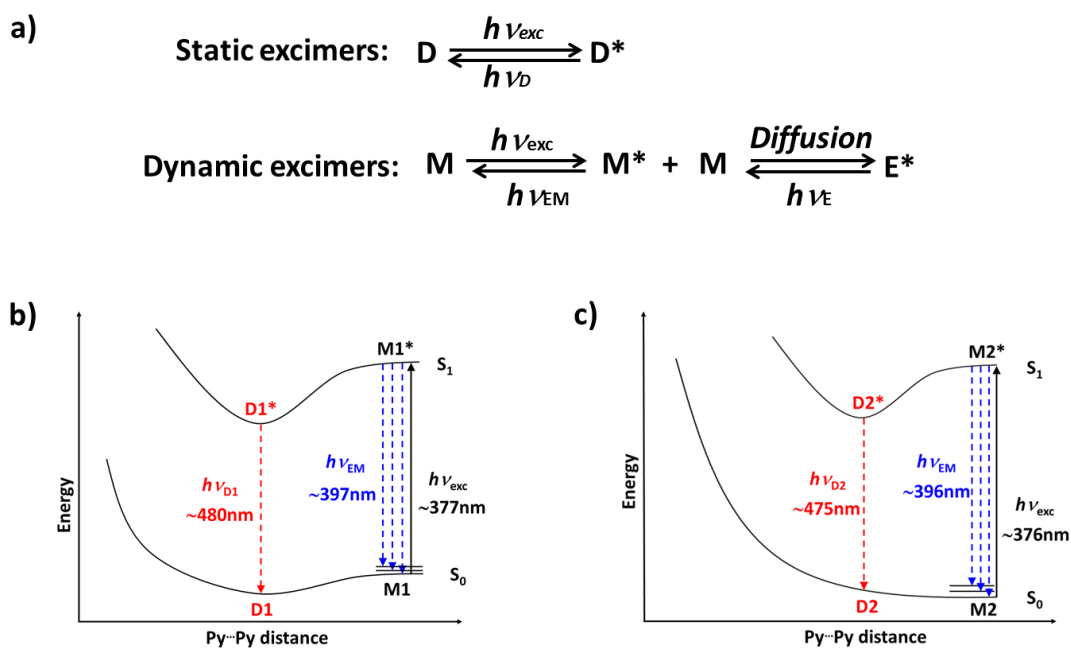

**Scheme 1.** a) Kinetic schemes of excimer formation in static and dynamic excimers. A hypothetical schematic potential energy surfaces for strongly- and weakly-coupled static excimers in b) PyMA1 and c) PyMA2, respectively. M1, M2 = Monomers with isolated electronic states in PyMA1 and PyMA2; D1 and D2 = Strongly- and weakly-coupled ground state dimers in PyMA1 and PyMA2, respectively. D1* and D2* = Static excimers with strong and weak excitonic coupling in PyMA1 and PyMA2, respectively.

Single crystal structures of PyMA1 and PyMA2 shed lights on the intermolecular interactions between the dimers and the extent of overlapping of the pyrene moieties in them, in solid state. The molecular conformations of PyMA1 and PyMA2 in single crystals at 80K are quite similar (Appendix Table S2 and S3). In both the crystal structures, N-H⋯O hydrogen bond and π⋯π interactions between the unit translated molecules are holding the entire packing of molecules and the distance and angle parameters suggest that the extent of the interactions are very similar in both the crystals (Figure 2a,b; Appendix Figure S7, S8, S9 and S10; Table S4 and S5). The approximate overlap ratios of the two pyrene moieties in the crystal structures, at 80K, are ~17% for PyMA1 and ~42% for PyMA2 (Figure 2c,d).

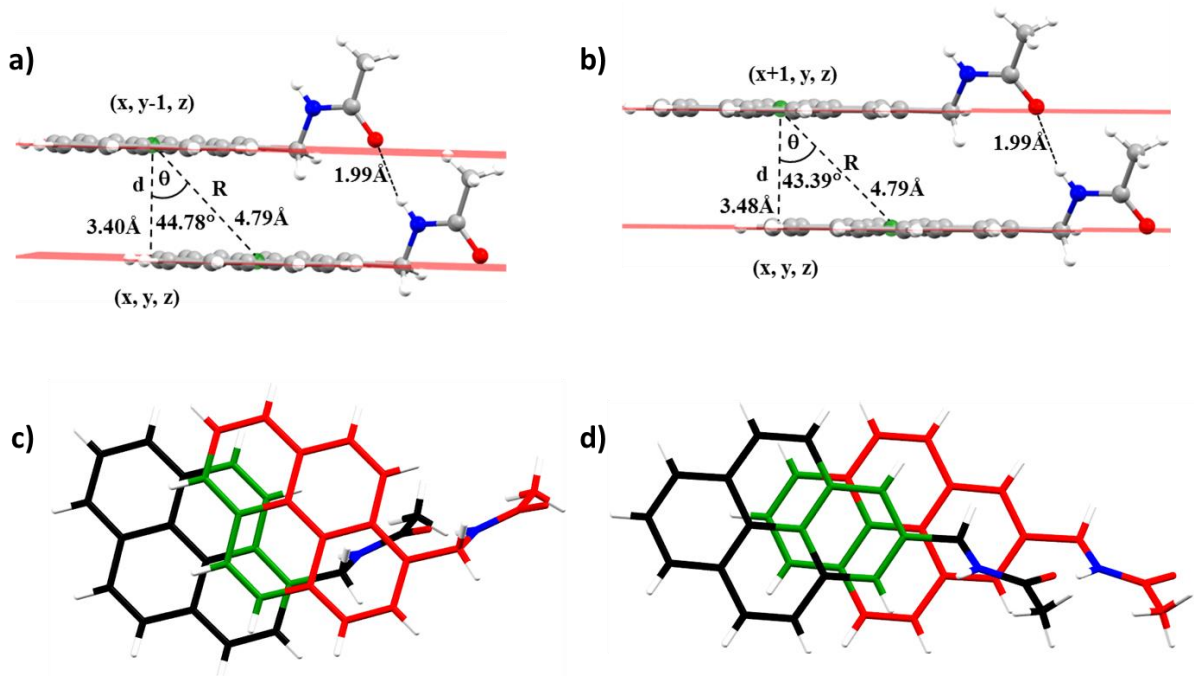

**Figure 2**. Parallel displaced π···π interaction and N-H···O hydrogen bond interaction between the unit translated molecules in a) PyMA1 and b) PyMA2. Centroids are shown in green. Extent of overlap between the pyrene rings of unit translated molecules, c) PyMA1 (~17%) and d) PyMA2 (~42%) are showed with green colored atoms. The carbon atoms of the unit translated molecules are coloured in red and black.

The time-correlated single photon counting (TCSPC) technique was employed to determine the decay lifetime of the emissive species for both PyMA1 and PyMA2 in solution and as single crystals. A pulsed 375nm laser operated at 4MHz was used to excite all the samples. In toluene, emissive species in PyMA1 and PyMA2 decay mono-exponentially with decay lifetimes in the range of ~1-3ns through all concentrations (Appendix Table S6; Figure S11). However, at 1.0mM PyMA1 shows a lifetime of ~29*ns* with small contribution (~0.02%), which can be identified as excimer emission. The lifetimes measured for both systems at 1.0mM are considerably longer than at 0.1mM (Appendix Table S6). In single crystals, both PyMA1 and PyMA2 decay bi-exponentially when excited at 375nm, suggesting the presence of both monomeric and excimer species. As the temperature was reduced slowly from 296K to 77K, the population of the excimer species was gradually increased in PyMA1 and PyMA2 crystals from 1% to 5% and 2% to 11%, respectively (Appendix Table S6 and Figure S11). The π-π overlap between the pyrene moieties belonging to the unit translated molecules in PyMA2 (~42%) is comparatively better than PyMA1 (~17%) at 80K and must have caused the higher lifetime as well as the percentage of excimers for the former.[27] While for PyMA1, the excimer lifetime systematically decreased from ~35.6*ns* to ~23.5*ns* with the gradual reduction of temperature, for PyMA2 the excimer lifetimes do not change much through the different temperatures.

The time-dependent DFT (TDDFT) gas phase calculations were employed to anticipate the electronic transition energies and intensities of singlet to singlet transitions for the individual molecules (Figure 3). Theoretically obtained absorption spectra for PyMA1 and PyMA2 molecules are in well agreement with the experimentally obtained spectra collected at 0.1mM in toluene (Figure 3a). The long wavelength absorption bands are slightly bathochromically shifted for PyMA1 compared to PyMA2, since the electron withdrawing ability of methylacetamide substituted on the pyrene ring is higher when substituted at the position 1.[28] The substitution at positions 2- and 7- makes these sites rather intractable for electrophilic substitution as the nodal plane in HOMO and LUMO in pyrene passes through them.[29] The HOMO and LUMO have contributed mostly in the excitation transitions for both PyMA1 and PyMA2 as shown in Figure 3b,c. The HOMO-LUMO gap for PyMA1 is slightly smaller than PyMA2 (~0.08 eV), clearly exhibits that substitution position 1 is more sensitive than position 2. The energy values and the frontier orbitals diagrams suggest HOMO to LUMO is a $\pi^* \leftarrow \pi$ transition for both, PyMA1 and PyMA2.

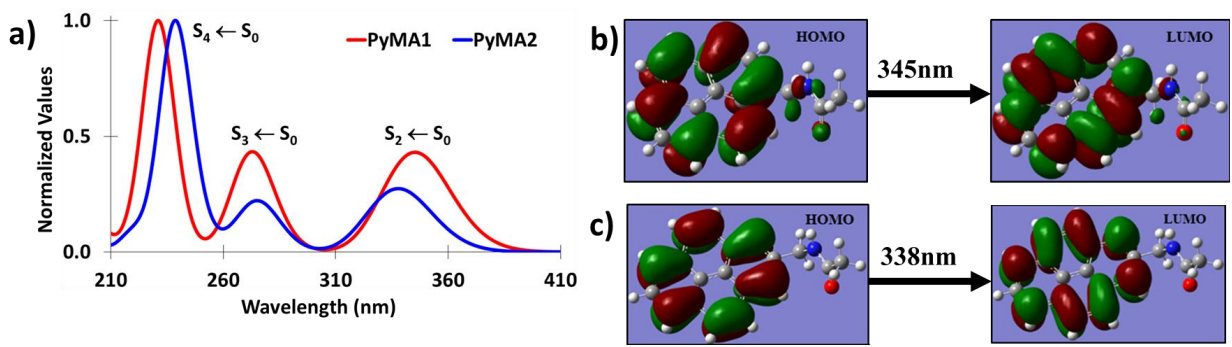

**Figure 3**. a) Theoretical absorption spectra of PyMA1 and PyMA2 molecules are calculated by TDDFT method at B3LYP/6-311G** level of theory. The molecular frontier orbitals, HOMO and LUMO in b) PyMA1 and c) PyMA2.

The excited state absorption of the transient species in PyMA1 and PyMA2 were investigated by ultrafast optical transient absorption (TA) spectroscopy. We have employed *ns*-TA (EOS mode) and *fs*-TA (HELIOS mode), to explore the excited state dynamics of the pyrene-*N*-methylacetamide isomers, in *ns-μs* and *fs-ps* time domains, respectively. PyMA1 and PyMA2 solutions at 1.0mM, in toluene, were excited by 345nm laser radiation at 200μW and subsequently probed with a weak white light continuum probe. The selection of excitation wavelength is very crucial to remove ambiguity while studying the kinetics and mechanism of excimers.[30] Prior to the measurements the solutions were purged with $N_2$ gas. For HELIOS mode, the continuum probe ranges between ~440-760nm whereas for EOS mode, a fibre coupled continuum laser with ~370-900nm spectrum is used as the probe. TDDFT calculations and spectroscopic measurements have suggested that the excitation with 345nm radiation will take PyMA1 and PyMA2 molecules into the excited state singlet $S_2$ through $S_2 \leftarrow S_0$ transition. Followed by the excitation into $S_2$, the monomers would rapidly go through a fast process of internal conversion (IC) into the $S_1$ state. The absence of any relevant negative signal (ground state bleach) could be attributed to the range of probe in TA spectra, which is outside the region of ground state absorption and also a high molar absorptivity for singlet – singlet ($S_N \leftarrow S_1$) and triplet – triplet ($T_N \leftarrow T_1$) absorption within the same range.

Figure 4a shows the formation of two major bands in the *fs*-TA spectrum of PyMA1, centred on 520nm and ~595nm, almost simultaneously (Inset: Figure 4a). The evolution of the absorption band around 595nm, in ~ 560*fs*, can be identified as the monomeric $S_2$ state of PyMA1. Similar sort of band for pyrene monomers is also commonly observed around ~585nm.[31] The decay of 595nm band in ~160*fs* by monomeric $S_1 \leftarrow S_2$ transition (IC) are in good agreement with the previously published results.[23,31] The evolution of the 520nm band which initially takes a form in ~560*fs*, can be entirely assigned to the absorption from static excimers. The formation time for static excimer observed here resembles the time scale proposed previously (~500*fs*) for excimer formation in pyrene crystals.[32] The broad 520nm absorption band completes its formation in ~(700-800)*fs* (Figure 4b), which coincides with the decay of $S_2$ and can be considered as an overlap of the absorption contributions from both monomeric $S_1$ and static excimers. Since, pyrene excimers have strong absorption bands in the wavelength region between 400nm – 550nm which overlaps with the $S_N \leftarrow S_1$ transitions.[23,25,33-35] It is also important to note, as exhibited in the Figure 4, rapid formation of a band around 480-510nm was also observed for both PyMA1 and PyMA2, and can be identified as the "coherent artifact" contribution from toluene, which completely decays within a few *ps* (~4*ps*) (Appendix Figure S15 and S16).

The *fs*-TA spectrum for PyMA2 shows the formation of two very broad bands ~530nm and ~580nm. The 580nm band which forms in ~630*fs*, gets bathochromically shifted to 600nm within ~1.2*ps*, can be identified as the monomeric singlet $S_2$ state (Figure 4c). The inception of the decay of the 580nm/600nm band is followed by the appearance of 530nm band, as the isosbestic point around 540-550nm showed. The 530nm absorption band completely forms within ~1.4*ps* and can be identified as monomeric $S_1$ state (Figure 5d).[23]

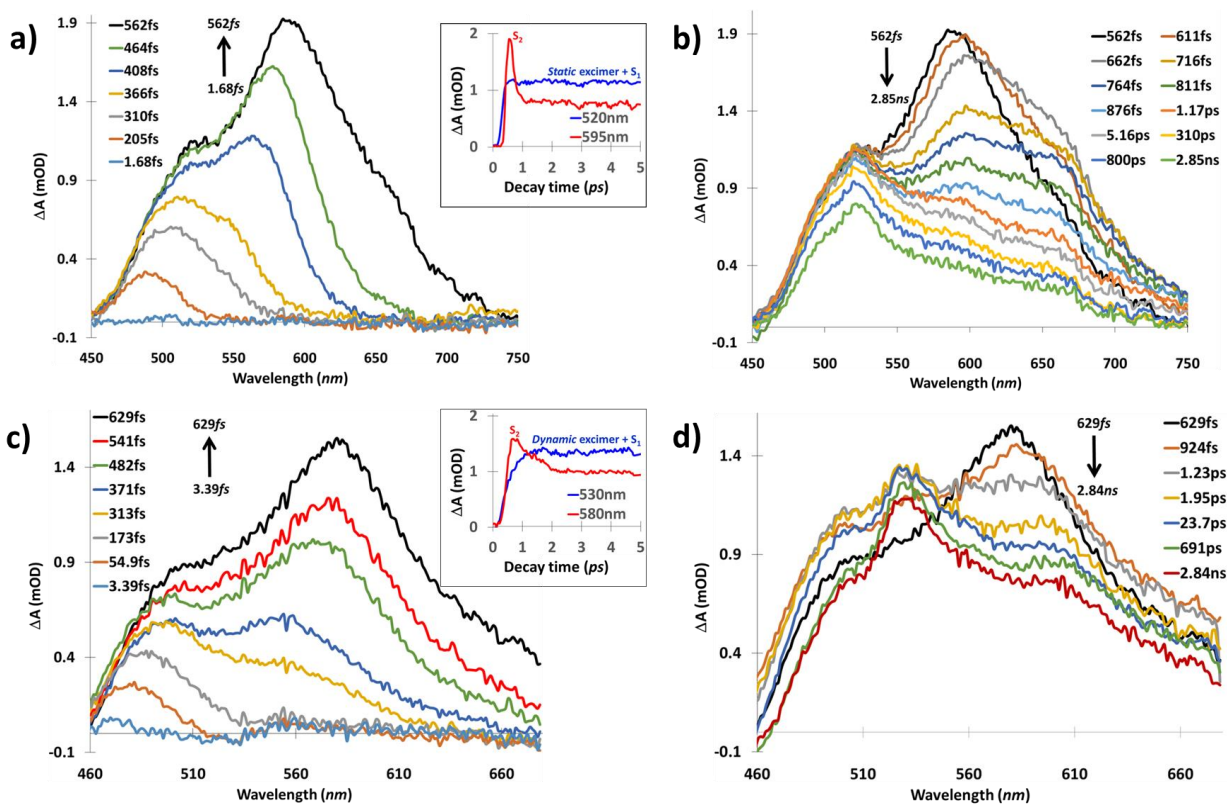

**Figure 4**. The *fs*-TA spectra at 1.0mM in toluene. PyMA1 (a, b). PyMA2 (c, d). Inset: Kinetic profile of a) 520nm and 595nm bands in PyMA1 and c) 530nm and 580nm bands in PyMA2. ($\lambda_{exc}$ = 345nm).

The correlation between the decay and rise of 580nm/600nm and 530nm bands, respectively, can be witnessed in their respective kinetic profiles while plotted together (Inset: Figure 4c). Noticeably, the decay of $S_2$ is much slower for PyMA2 (Inset: Figure 4c). The appearance of the 530nm absorption band with the prominent shoulder around ~500nm could be attributed to the overlapping contributions originated from monomeric and excimer absorptions. The kinetic profile corresponding to either the 500nm or 530nm on the *fs*-TA spectra shows that the excimer formation between weakly-coupled pre-associated dimers must have completed by ~65*ps*, as both bands start decaying only after that time point. Since, ~65*ps* would be considered extremely fast for monomeric fluorescence, the two other possible ways that the monomeric excited singlet states could start relaxing are either by ISC or by the formation of static excimers between weakly-coupled pre-associated dimers after a simple reorientation or both.

The singular value decomposition (SVD) technique, involving 3-dimensional ΔA vs time and wavelength spectroscopic data, was employed. It was followed by global analyses to find the principal components responsible for transient signals observed in *fs*-TA and *ns*-TA datasets in the form of decay associated spectra (DAS) and the kinetic profile of the decaying components. The global analyses on *fs*-TA data collected with PyMA1 shows principle components represented by three right singular vectors in the DAS. The first vector (V1) decays tri-exponentially, with the major species, $\tau_1$ (98%), affiliated with the DAS centred on 580nm, identified as $S_1 \leftarrow S_2$ IC. Two other minor species, $\tau_2$ and $\tau_3$ represent two slower processes, which could be $S_0 \leftarrow S_2$ fluorescence[36,37] or $T_2 \leftarrow S_2$ intersystem crossing (ISC)[38] (Appendix Table S9; Figure S9a). Whereas the second component (V2) centred on 520nm and the third component (V3) containing bands centred on 520nm, 600nm and 660nm have decayed only partially (Appendix Table S9; Figure S9a).

The global analyses suggest that the entire *fs*-TA spectrum for PyMA2 can be associated with three singular vector components, centred on 580nm (V1), 530nm (V2) and 485nm (V3). The decay profile of the right singular vector V1 from SVD analysis shows that about 93% of the molecules in the $S_2$ state depopulate to the $S_1$ state through IC. While the rest of the population decays through slower processes like $S_0 \leftarrow S_2$ transition or $T_2 \leftarrow S_2$ ISC. The other two components centred on 530nm (V2) and 485nm (V3) have decayed only partially (Appendix Figure S9b).

The EOS temporal mode was employed to capture the excited state dynamics in the *ns-μs* regime with 100*ps* time resolution. The *ns*-TA data sets for PyMA1 and PyMA2 at 1.0mM were collected from 12*ns* till 2*μs*. The *ns*-TA spectra of PyMA1 and PyMA2 are quite similar in nature (Figure 5). In the beginning, the spectrum for PyMA1 shows a broad band centred on 490nm, with prominent shoulder at 520nm, which starts decaying after ~12.6*ns* and the 520nm shoulder takes the shape of a distinct band, well separated from the 490nm band. The relative intensity of the 520nm band had also increased with respect to the 490nm band, with time (Figure 5a). For PyMA2 the 490nm and 530nm absorption bands were well separated from the beginning and started decaying after ~12.7*ns*. For PyMA2 also, while decaying, the relative intensity of the 530nm absorption band had increased with respect to the 490nm band. The 520nm and 530nm absorption bands for PyMA1 and PyMA2 respectively can be identified with the overlapping of absorption contributions originated from monomeric triplet and excimer singlet. For both the systems, the decay of the 490nm band is juxtaposed with the rise of the 425-428nm band which forms in ~40*ns* and ~50*ns* for PyMA1 and PyMA2, respectively. The 490nm and 425-428nm absorption bands could be identified mostly from monomeric singlet and triplet respectively, as suggested by the isosbestic point around 450-460nm region in the *ns*-TA spectra as well as their relative kinetic profiles (Appendix Figure S10). The global analyses of the *ns*-TA spectrum for PyMA1 suggest the presence of two components or right singular vectors. The right singular vector V1 at 480nm has two components; the majority (~75%) decays through $S_0 \leftarrow S_1$ monomeric transition, while, the second component (~25%) that decays in ~255*ns* can be identified as a monomeric triplet (Appendix Table S10). The reduction of triplet lifetime, at least by an order, could be due to the presence of residual oxygen in the solution.

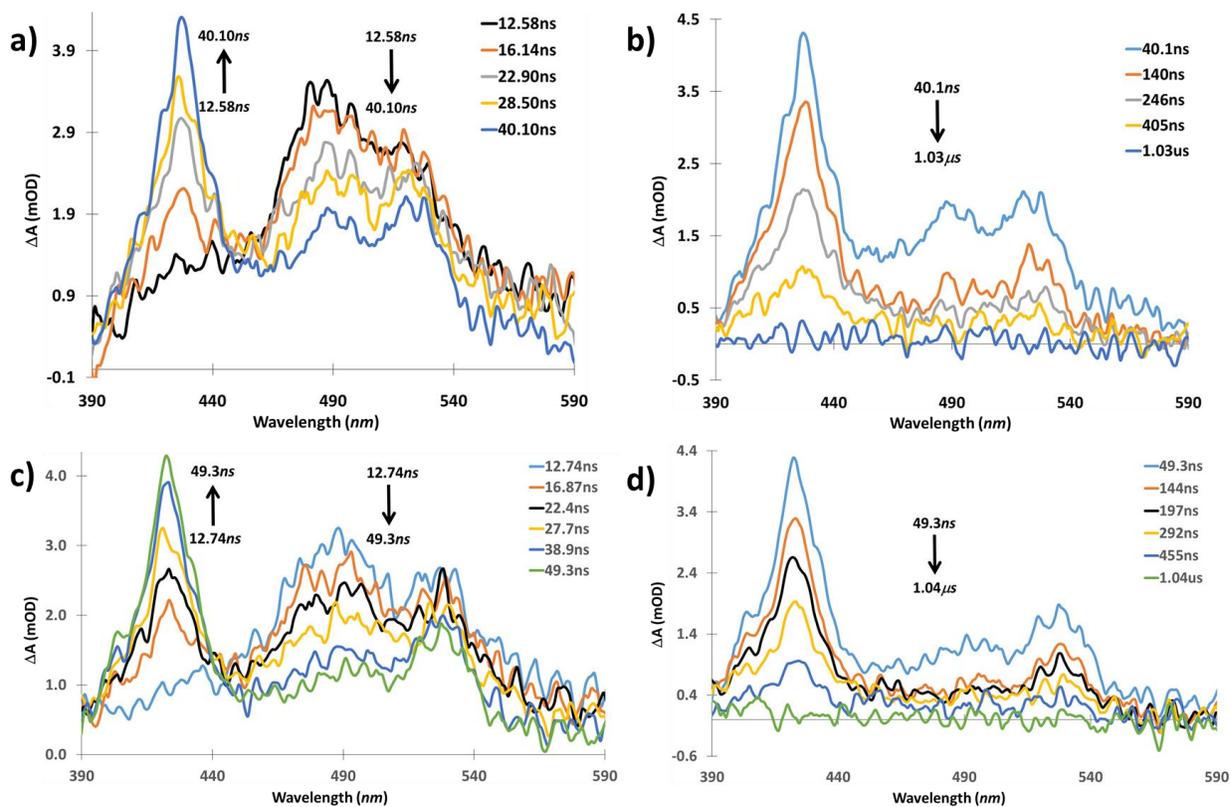

**Figure 5**. The *ns*-TA spectra of PyMA1 (a, b) and PyMA2 (c, d) at 1.0mM, in toluene. ($\lambda_{exc}$ = 345nm).

The right singular vector V2 has two components centred on 425nm and 525nm. While the component at 425nm decays mono-exponentially with lifetime ~239*ns*, the 525nm component decays bi-exponentially with lifetime ~28*ns* and ~240*ns*. While the longer lifetime in 525nm band could be identified as the monomeric triplet, the shorter lifetime could be identified as excimer singlet. We have performed the *ns*-TA measurements with PyMA1 at 0.1mM (data not presented here) which suggest a shorter lifetime of ~18*ns* in the decay component of the vectors, V2 at 525nm, in the absence of any excimers. Therefore it could be argued that the lifetime of ~28*ns* which consists of ~28% of the decay component of right singular vector V2 at 525nm is due to the absorption contribution from static excimers.

For PyMA2 also the global analyses of the *ns*-TA data suggest two right singular vectors that represent all the transient absorption signals. The right singular vector V1 at 480nm decays bi-exponentially with 78% decaying through $S_0 \leftarrow S_1$ monomeric transition and rest of the 22% that decays in ~263*ns* can be identified as monomeric triplets. As observed for PyMA1, the right singular vector V2 at 425nm for PyMA2 represents monomeric triplets with a lifetime of ~258*ns*. Decay of the right singular vector V2 at 530nm for PyMA2 exactly resembles the decay of the right singular vector V2 at 525nm for PyMA1. The right singular vector V2 at 530nm in PyMA2 decays bi-exponentially in ~21*ns* and ~253*ns*. The global analysis of the *ns*-TA data with PyMA2 at 0.1mM reveals a much shorter decay times for the component V2 at 530nm with ~6*ns*, in the absence of any excimers (data not presented here).

Excimer formation by intermolecular hydrogen bonds is not studied much in the context of pyrene derivatives. Due to its planar structure pyrene is already known to be susceptible to π⋯π stacking, essential for excimer formation. In fact, the need of functionalizing pyrene comes from the idea of getting rid of the excimer formation by π⋯π stacking. In that regard, the serendipitous encounter of strongly- and weakly-coupled static excimers for two isomeric mono-substituted pyrene derivatives where *N*-methylacetamide is substituted on pyrene at two different positions is quite interesting. This also encourages in exploring the possibility of designing different excimer types in pyrene based systems. The excitation wavelength of 345nm proved to be crucial for excimer formation in this case. While exciting the systems with 375nm; the formation of excimers is not so evident in solutions but quite noticeable in the single crystals. It could be that in solution, it was not possible to excite the 375nm band of the pre-associated dimers as at this excitation wavelength they are a minority component compared to the monomers. The intermolecular hydrogen bonds may have contributed in the absence of multiple clusters of molecules with different degrees of overlap in the same solution during dimerization, as reflected in the singular decay time of excimers for both the systems.[39] The results obtained from the *fs*-TA measurements suggest that the strongly-coupled pre-associated dimers in the ground state of PyMA1 formed the static excimers, in ~560*fs*, upon photoexcitation and are proved to be a much faster process than the formation of static excimers with weakly-coupled pre-associated dimers in PyMA2, which in this case most likely formed within ~65*ps*. Additional time may have needed for the weakly-coupled pre-associated dimers to reorient itself to form excimers. It is also observed that the strongly- and weakly-coupled static excimers have quite a different formation time but their respective decay lifetimes with ~28*ns* and 21*ns*, respectively, as showed by *ns*-TA, are not so different. The excimer lifetimes obtained from TCSPC study with single crystals closely matches with the results obtained from TA measurements performed in solutions. In future, it would be interesting to employ time-resolved photocrystallography measurements to capture the excimer geometry in single crystals.[40,41] Lately, several examples of excimer induced enhanced emission (EIEE) are reported, which encourages excimer formations to be well utilized in designing multifunctional optoelectronic materials.[42-44] In that regard, pyrene based systems are better suited than other conjugated systems due to its range and easy tunability of its photo-physical properties by substituting different functional groups at different positions. Further studies in the quest of having a better understanding of the ultrafast processes like excimer dynamics is highly desired, to design novel multifunctional materials with pyrene based systems.

**Notes**

The authors declare no competing financial interests.

ACKNOWLEDGMENT
S.T. is grateful for financial support within the ECRAPS innovation fund of the HGF. Use of the Center for Nanoscale Materials, an Office of Science user facility, was supported by the U.S. Department of Energy, Office of Science, Office of Basic Energy Sciences, under Contract No. DE-AC02-06CH11357. A part of the research was carried out at the light source PETRA-III at DESY, a member of the Helmholtz Association (HGF). We would like to thank P11 staff for assistance. K.B. is grateful to Mr. Marten Rittner from Institute for Nanostructure and Solid State Physics for helping with the measurement of fluorescence emission spectra. The current work has been funded by the Deutsche Forschungsgemeinschaft (DFG, German Research Foundation) - 217133147/SFB 1073. K.B. acknowledges Dr. Sanjoy K. Mahatha for proof reading and technical feedback in preparing figures.

# Appendix

**Experimental and Theoretical Methods:**

*Synthetic Procedure.* PyMA1 and PyMA2 were synthesized adopting the procedures mentioned in the literature.[1-4]

*UV-Vis Absorption Spectral Measurement.* UV-Vis absorption spectra were collected using a Cary-5E UV-VIS spectrophotometer (Varian Australia).[5] The wavelength interval was 0.5nm and the path length of the beam inside the cuvettes were 1mm and 10mm. The absorption spectra for the compound of interest were corrected using a reference spectrum corresponding to the solvent that is used to dissolve the compound. All the measurements were carried out at RT.

*Fluorescence Emission and Excitation Spectra.* Jobin Yvon Horiba Model Fluorolog 3 FL3 22 equipped with, both, front-face (22°) and right angle (90°) detection was used to collect the fluorescence emission and excitation spectra. The instrument is also equipped with 450 W Xenon lamp for excitation. The measurements were performed using 3mm and 10mm path length quartz cuvettes. All the spectra were corrected using the correction files available in the Horiba software that deals with the excitation light intensity and photomultiplier (PMT) response. While collecting the fluorescence emission spectra the width of the entrance and exit slit width was 2*nm* for both emission and excitation. In case of collecting fluorescence excitation spectra, for PyMA1, the entrance and exit slit width for excitation was 2*nm* and for emission it was 1*nm*. While for PyMA2, the entrance and exit slit width for excitation was 3*nm* and for emission it was 1*nm*. The measurements in the solutions were performed after purging with $N_2$ gas for 15-20 minutes. We have also used Agilent Cary Eclipse Fluorescence Spectrophotometer to measure some of the spectra.

*X-ray Diffraction.* Suitable single crystals of PyMA1 and PyMA2 were grown in ethyl acetate/ethanol mixture by slow evaporation. X-ray data were collected on undulator synchrotron radiation with λ = 0.62073 Å at P11 beamline in PETRA III, DESY, Hamburg, Germany. Indexing of the X-ray diffraction pattern, unit cell refinement and spot integration were performed with XDS.[6] The crystal structures were solved and subsequently refined using the X-ray diffraction datasets collected at 80K for both the compounds. All the X-ray diffraction data sets were collected in phi scan type mode. For both the compounds, the crystal structure was solved using direct methods in SHELXS.[7] All the structures were refined against F2 isotropically, followed by full matrix anisotropic least-squares refinement using SHELXL-97.[8] For both the structures, all the hydrogen atoms were fixed geometrically, in idealized positions, and allowed to ride with the respective C or N atoms to which each was bonded, in the final cycles of refinement. CCDC deposition numbers for the compounds are 1896002 (PyMA1) and 1896013 (PyMA2) which contain the supplementary crystallographic data for this paper and can be obtained free of charge from The Cambridge Crystallographic Data Centre via www.ccdc.cam.ac.uk/data_request/cif.

*Time-correlated single photon counting.* Photoluminescence (PL) spectra and lifetimes were measured at the Center for Nanoscale Materials (CNM) at Argonne National Laboratory, using a home-made fluorescence microscope fitted with a liquid nitrogen cooled continuous flow cryostat (Janis ST-500UC) in CNM. The instrument was based on an Olympus IX-71 inverted microscope. A pulsed 375nm laser (PicoQuant, DC375M) operated at 4MHz was used to excite the sample through a ThorLabs LMU-15X-NUVobjective that was used to both focus the incoming laser light and collect the emitted PL. The collected PL was separated from the exciting laser using a dichroic mirror and a bandpass filter (both Semrock). The PL was then routed either to a spectrograph (Princeton Instruments, SpectraPro-300) fitted with a CCD camera (Princeton Instruments, PIXIS) or, for lifetime measurements, to a fiber-coupled single photon avalanche diode (SPAD) (Micro Photon Devices, PDM). The output from the SPAD and a trigger pulse from the laser power supply were fed to the two input channels of a time-correlated single photon counting (TCSPC) system (PicoQuant, PicoHarp 300).

*Quantum Chemical Calulations*. All the quantum chemical gas phase calculations were performed using DFT methods at B3LYP/ 6-311G** level of theory with the Becke[9] three-parameter hybrid functional and Lee−Yang−Parr's[10] gradient-corrected correlation functional (B3LYP) implemented in the Gaussian16 (G16)[11] package. The ground state optimized singlet was used for the calculation of frontier molecular orbitals (MOs) at occupied ground state and unoccupied virtual state by time dependent DFT (TDDFT) methods. TDDFT calculations have also provided theoretical UV-Vis spectra along with the excitation energy of the molecules at the gas phase. GaussSum 3.0 was used to plot the density of states diagram.[12] GaussView 6.0 was used to plot the frontier orbitals.[13] Due to the occurrence of many closely lying states within a small energy gap, the density of states (DOS) plot will help us to realize the total concentration of available states within a small given energy range (Figure S6). Although, already provided frontier orbital diagrams of HOMO and LUMO is showing that HOMO and LUMO are mostly occupied by pyrene moiety, nonetheless, the DOS plot provides a more clear and quantitative picture with the exact values of energy ranges.

*Optical Transient Absorption Spectroscopy*. Ultrafast transient absorption (TA) spectra and kinetics were carried out at the Center for Nanoscale Materials (CNM) at Argonne National Laboratory using an amplified Ti:sapphire laser system (Spectra Physics, Spitfire-Pro) and automated data acquisition system (Ultrafast Systems, Helios). The amplifier produced 120fs pulses at 5kHz. The output from the amplifier was split 90/10 to pump an optical parametric amplifier (TOPAS) and to generate a continuum (450-750nm) probe. The TA system enables three-dimensional data collection (spectra/time/ΔOD). The continuum probe beam was generated by first sending the 10% output from the laser amplifier down a computer controlled optical delay line then focusing into a 3mm thick piece of sapphire. The residual 800nm light was removed from the probe beam with an interference filter. The probe beam was then focused using a 20X microscope objective onto a horizontally held 2mm quartz cuvette containing a solution. The transmitted probe beam was detected using a fiber optically coupled spectrograph with a 1D, 2048-pixel CCD array detector. The excitation beam was overlapped on the probe beam spot on the sample at an incident angle of ca. 15 degrees after being optically chopped at 2.5kHz using a synchronous chopper so that the spectrograph measured the transmitted probe beam alternatively as $T_{ON}$ and $T_{OFF}$. The differential extinction $\Delta A= \log_{10}(T_{ON}/T_{OFF})$ was calculated for each pair of pulses and was typically averaged over a two second interval for each delay time. Temporal chirp in the probe pulse was measured and corrected for by making a measurement on neat solvent; the resonant signal was then fitted for each probe wavelength to determine the zero-delay position between pump and probe. For the measurements, typical experimental excitation parameters were: 200 μW at 345nm. The probe beam is focussed to typically ~150-200μm spot. It is necessary to make sure that no area is getting probed which is not pumped first. The pump beam size is always slightly bigger than the probe beam since there is some spatial chirp on the probe beam. In both the temporal mode of data collection, HELIOS (0-3*ns*) and EOS (0-400*μs*), the pump beam is the same the only difference is the probe beam. The probe is generated differently for both the modes. For HELIOS mode, Sapphire (~440-760) crystal is used and for EOS mode, a fibre coupled continuum laser with ~370-900nm spectrum is used. For *fs-ps* regime, in the HELIOS mode, the excited state absorption was measured between -1.00*ps* to 2.84*ns*. First 1ps with 100*fs* step size, next 5*ps* with 50*fs* step size, next 500*ps* with 5*ps* step size and finally 2336*ps* with 20*ps* step size. The smallest step size was kept between 0 to 5*ps* with 50*fs* as most of the ultrafast processes were expected to observe in this time interval. In EOS mode the temporal resolution was 100*ps* throughout the measurement range. The measurements in the solutions were done after purging with $N_2$ gas for 15-20 minutes. All the solutions were continuously stirred with the help of a rotating magnet, while placing a small magnetic stir bar inside the cuvette. The magnet was rotating at 650rpm. TA datasets were processed using Surface Xplorer software provided by Ultrafast Systems.

**Table S1.** Crystal data and Structure refinement parameters of N-(pyren-1-ylmethyl) acetamide (**PyMA1**) and N-(pyren-2-ylmethyl) acetamide (**PyMA2**).

| Compounds | PyMA1 | PyMA2 |
|---|---|---|
| Empirical formula | C19 H15 N1 O1 | C19 H15 N1 O1 |
| Crystal color/habit | Colourless/ Needle | Colourless/ Needle |
| Crystal size (mm) | (0.50 x 0.49 x 0.28) | (0.60 x 0.08 x 0.02) |
| Crystallizing solvent | Ethyl Acetate / Ethanol | Ethyl Acetate / Ethanol |
| Crystal system/Space group | Orthorhombic / $Pna2_1$ | Triclinic / $P$-1 |
| $a$ (Å) | 22.580(3) | 4.7886(15) |
| $b$ (Å) | 4.790(4) | 11.9687(18) |
| $c$ (Å) | 12.4300(10) | 11.990(3) |
| $\alpha$ (°) | | 97.862(3) |
| $\beta$ (°) | | 95.497(4) |
| $\gamma$ (°) | | 99.642(7) |
| Volume (Å$^3$) | 1344.4(10) | 666.1(3) |
| Z/Z′ | 4/1 | 2/1 |
| Molecular weight | 273.32 | 273.32 |
| Calculated density (g/cm$^3$) | 1.350 | 1.363 |
| F (000) | 576 | 288 |
| Radiation | Synchrotron ($\lambda$ = 0.62073 Å) | Synchrotron ($\lambda$ = 0.62073 Å) |
| Temperature (K) | 80(2) | 80(2) |
| θ range (°) | 1.58-22.83 | 1.509-26.681 |
| Scan type | ϕ | ϕ |
| Measured reflections | 17158 | 25941 |
| Unique reflections | 1425 | 3856 |
| Observed reflection [ \|F\| > 4σ(F)] | 1422 | 3717 |
| Final R (%) | 3.17 | 5.57 |
| $w$R2 (%) | 8.25 | 17.96 |
| Goodness-of-fit on F$^2$ (S) | 1.055 | 1.092 |
| Δρ max (e.Å$^{-3}$) | 0.217 | 0.563 |
| Δρ min (e.Å$^{-3}$) | -0.237 | -0.433 |
| No. of restraints/parameters | 1/192 | 0/191 |

| Data[│F│ > 4σ(F)]-to-parameter ratio | 7.41: 1 | 19.46:1 |

**Table S2**. Bond lengths and bond angles of the *N*-methylacetamide moieties in single crystal structures of **PyMA1** and **PyMA2**.

| Bond Lengths (Å) | | | | | Bond Angles (°) | | | | |
|---|---|---|---|---|---|---|---|---|---|
| C6-C5 | C5-N4 | N4-C2 | C2-O3 | C2-C1 | C6-C5-N4 | C5-N4-C2 | N4-C2-C1 | N4-C2-O3 | O3-C2-C1 |
| **PyMA1** | | | | | | | | | |
| 1.513 | 1.463 | 1.338 | 1.234 | 1.507 | 112.10 | 120.76 | 117.03 | 122.22 | 120.74 |
| **PyMA2** | | | | | | | | | |
| 1.511 | 1.462 | 1.343 | 1.235 | 1.510 | 111.78 | 120.35 | 116.48 | 122.68 | 120.84 |

**Table S3**. Two major backbone torsion angles involving the *N*-methylacetamide moieties of **PyMA1** and **PyMA2**.

| Torsion angles (°) | | Inter-planar angles and centroid-to-centroid distances between pyrene ring and N-methylacetamide moiety | |
|---|---|---|---|
| α (C6-C5-N4-C2) | β (C5-N4-C2-C1) | Inter-planar angle (°) | centroid-to-centroid distance (Å) |
| **PyMA1** | | **PyMA2** | |
| 168.12 | 179.89 | 78.52 | 5.93 |
| N-(pyren-2-ylmethyl) acetamide (**2**) | | N-(pyren-2-ylmethyl) acetamide (**2**) | |
| 166.61 | 177.71 | 64.54 | 6.83 |

**Table S4**. Distance and angle parameters for hydrogen bonds and C-H···π interactions in **PyMA1** and **PyMA2** are listed.

| Donor | Acceptor | D···A(Å) | H···A(Å) | D-H···A(°) |
|---|---|---|---|---|
| **PyMA1** | | | | |
| N-H···O hydrogen bond | | | | |
| N4 (x, y, z) | O3 (x, y-1, z) | 2.86 | 1.99 | 167.84 |
| C-H···π interaction | | | | |
| C5 (x, y, z) | C20 (pyrene ring) (x, y+1, z) | 3.53 | 2.75 | 136.22 |
| C5 (x, y, z) | C7 (pyrene ring) (x, y+1, z) | 3.71 | 2.81 | 150.37 |
| C12 (pyrene ring) (x, y, z) | C16 (pyrene ring) (-x+1/2, y-1/2, z-1/2) | 3.57 | 2.78 | 141.41 |
| **PyMA2** | | | | |
| N-H⋯O hydrogen bond | | | | |
| N4 (x, y, z) | O3 (x+1, y, z) | 2.83 | 1.99 | 159.04 |
| C-H⋯π interaction | | | | |
| C5 (x, y, z) | C8 (pyrene ring) (x-1, y, z) | 3.62 | 2.75 | 146.09 |
| C5 (x, y, z) | C7 (pyrene ring) (x-1, y, z) | 3.75 | 2.77 | 168.46 |

**Table S5.** Distance and angle parameters for parallel displaced π⋯π interactions.

| Centroid-centroid distance R (Å) | Inter-planner angle (°) | Normal to the second π-plane from the centroid of the first π-system, d (Å) | θ (°) |
|---|---|---|---|
| **PyMA1** | | | |
| π⋯π interactions between unit translated molecules along '*b*' axis | | | |
| 4.79 | 0 | 3.40 | 44.78 |
| **PyMA2** | | | |
| π⋯π interactions between unit translated molecules along '*a*' axis | | | |
| 4.79 | 0 | 3.48 | 43.39 |

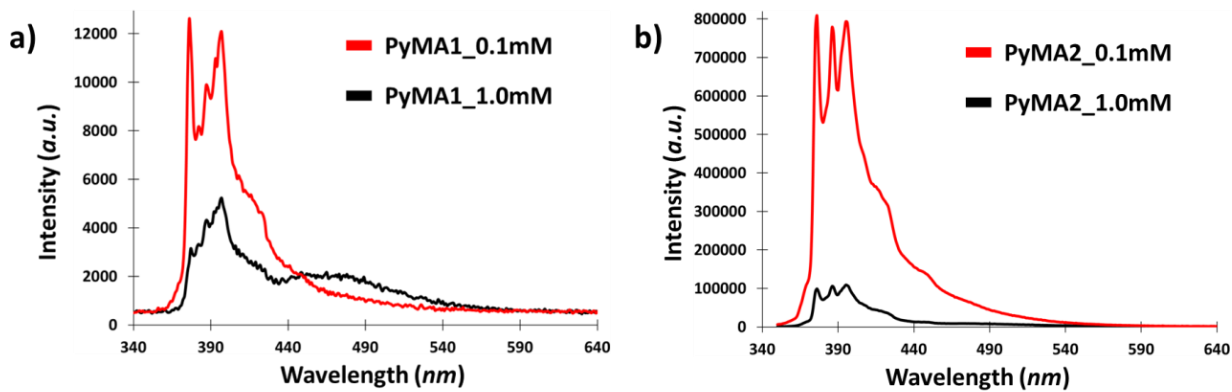

**Figure S1.** Emission spectra of, a) PyMA1 and b) PyMA2, at different concentrations in toluene (with 10mm cuvette).

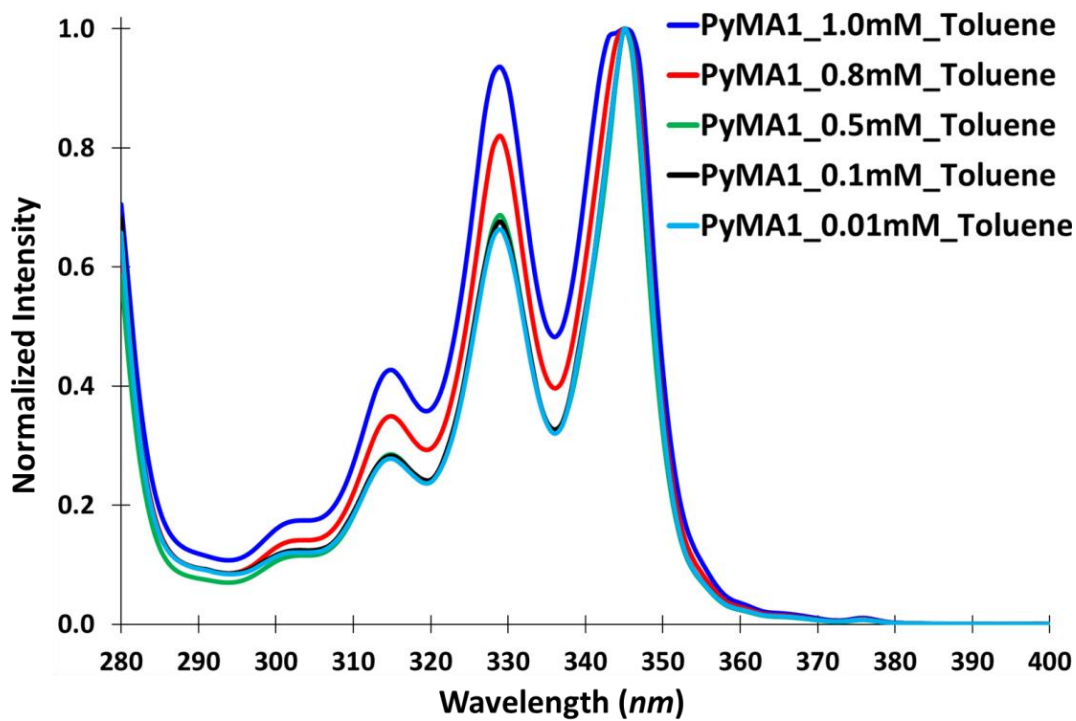

**Figure S2.** UV-Vis absorption spectra of PyMA1, at different concentrations in toluene (with 1mm cuvette).

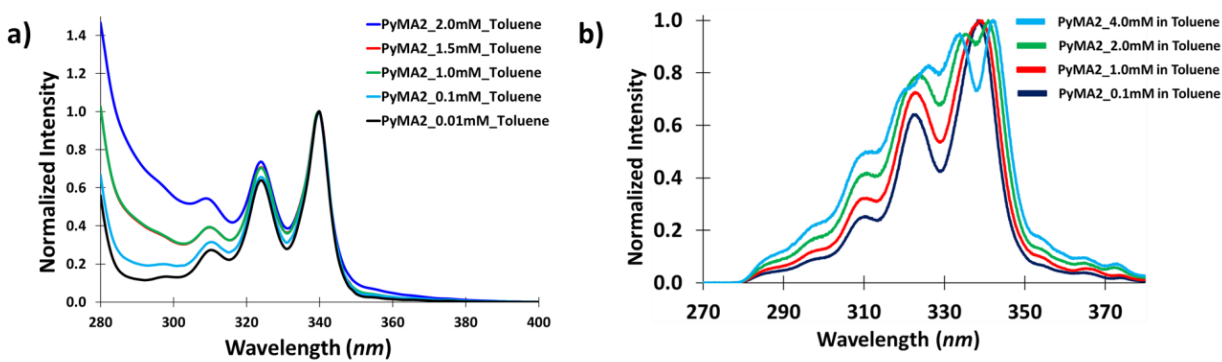

**Figure S3.** a) UV-Vis absorption spectra (with 1mm cuvette) and b) excitation spectra (with 10mm cuvette) of PyMA2, at different concentrations in toluene.

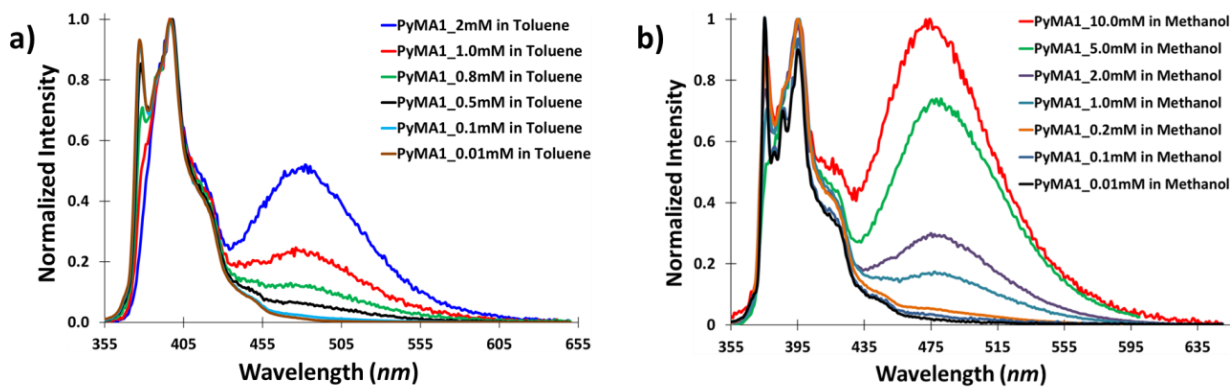

**Figure S4.** Emission spectra of PyMA1, at different concentrations in a) toluene and b) methanol (with 10mm cuvette).

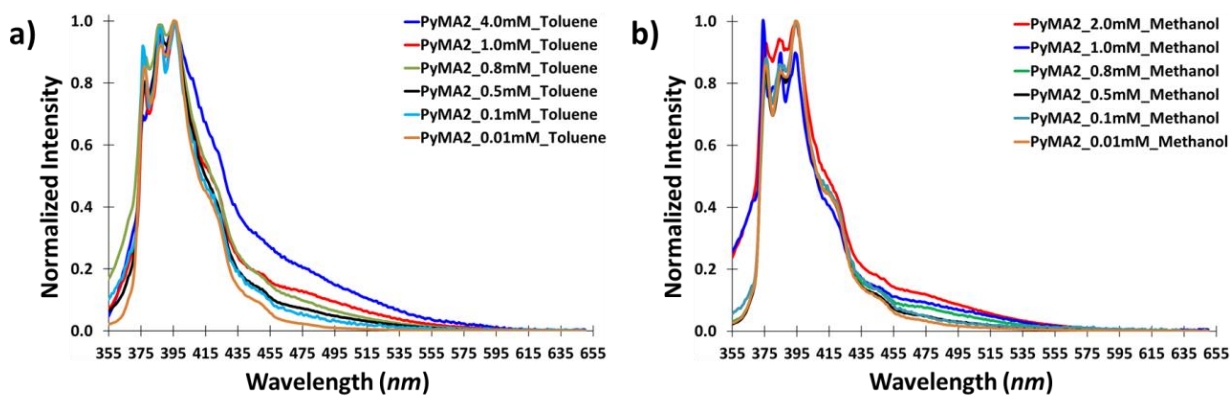

**Figure S5.** Emission spectra of PyMA2, at different concentrations in a) toluene and b) methanol (with 3mm cuvette).

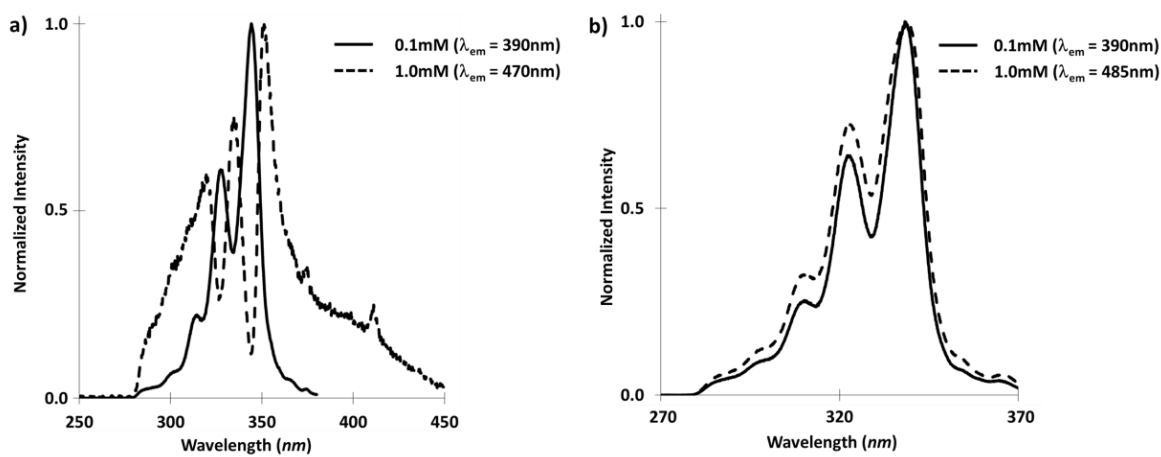

**Figure S6.** Normalized excitation spectra of, a) PyMA1 and b) PyMA2, at 0.1mM (solid lines) and 1.0mM (dashed lines) (with 10mm cuvette).

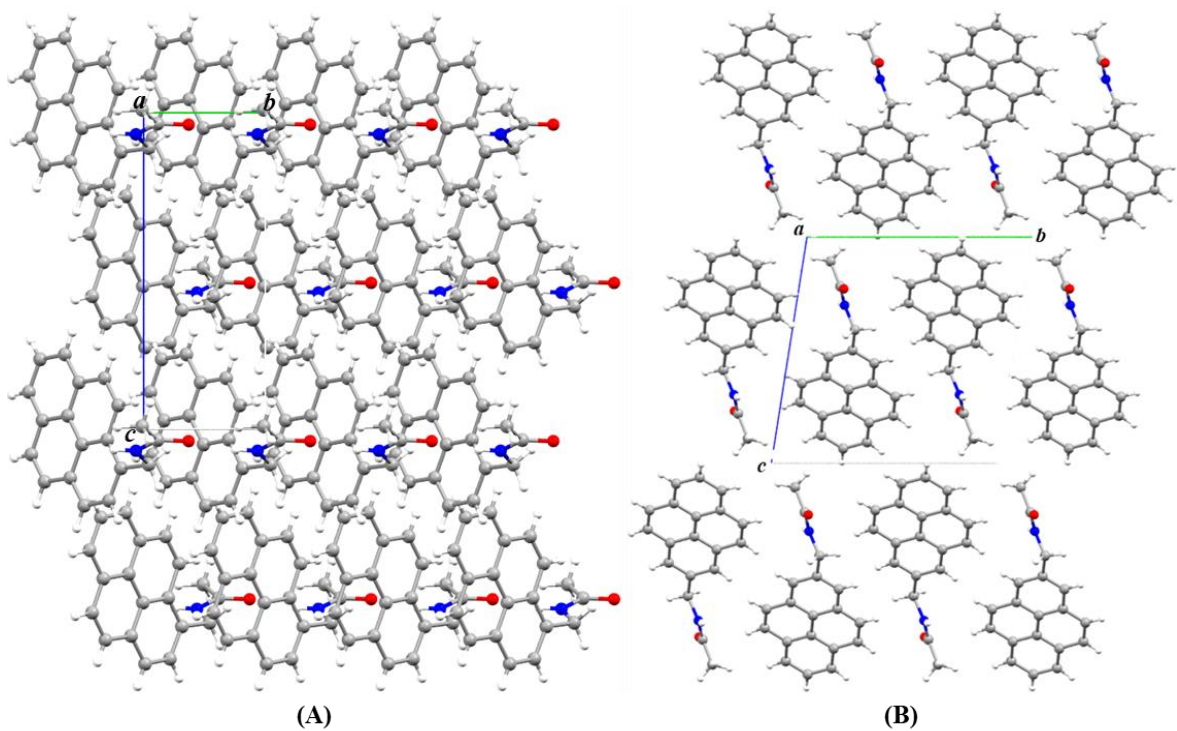

**Figure S7**. Packing of molecules down crystallographic '*a*' axis. A) **PyMA1**. B) **PyMA2**.

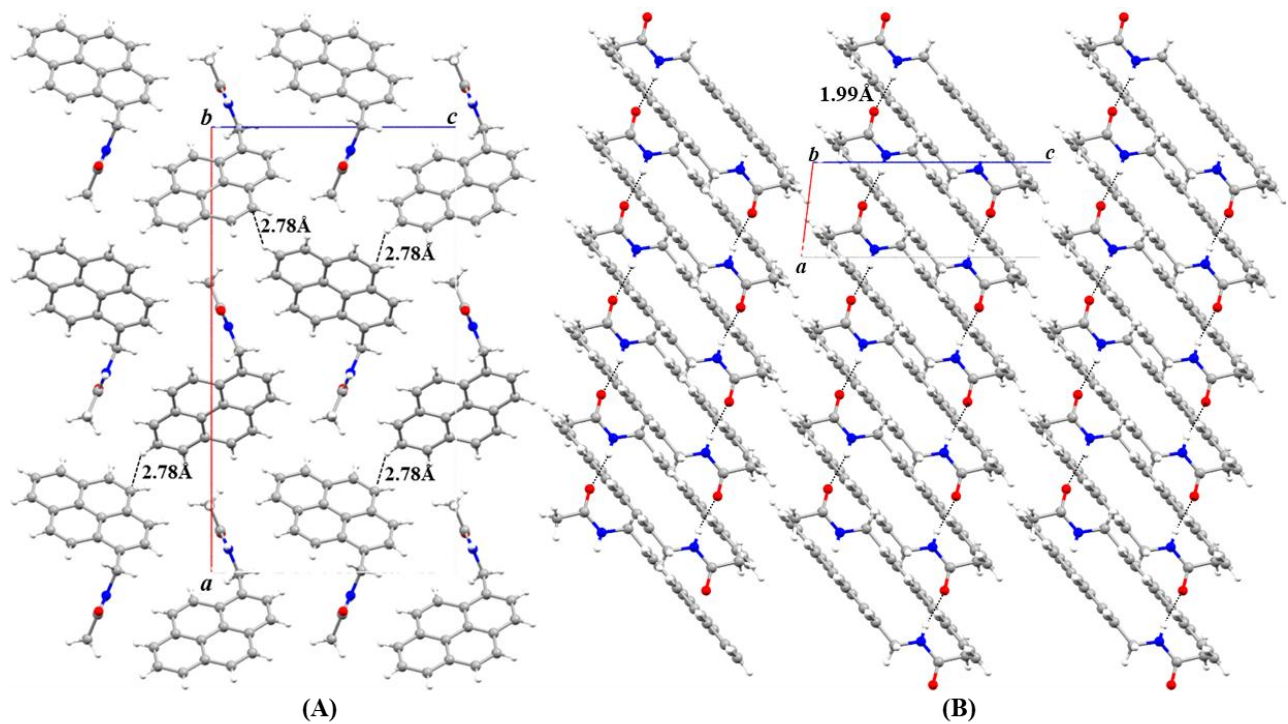

**Figure S8**. Packing of molecules down crystallographic '*b*' axis. A) **PyMA1** (C-H···π contacts are noted on the figure). B) **PyMA2** (H···O distance in N-H···O hydrogen bond is provided).

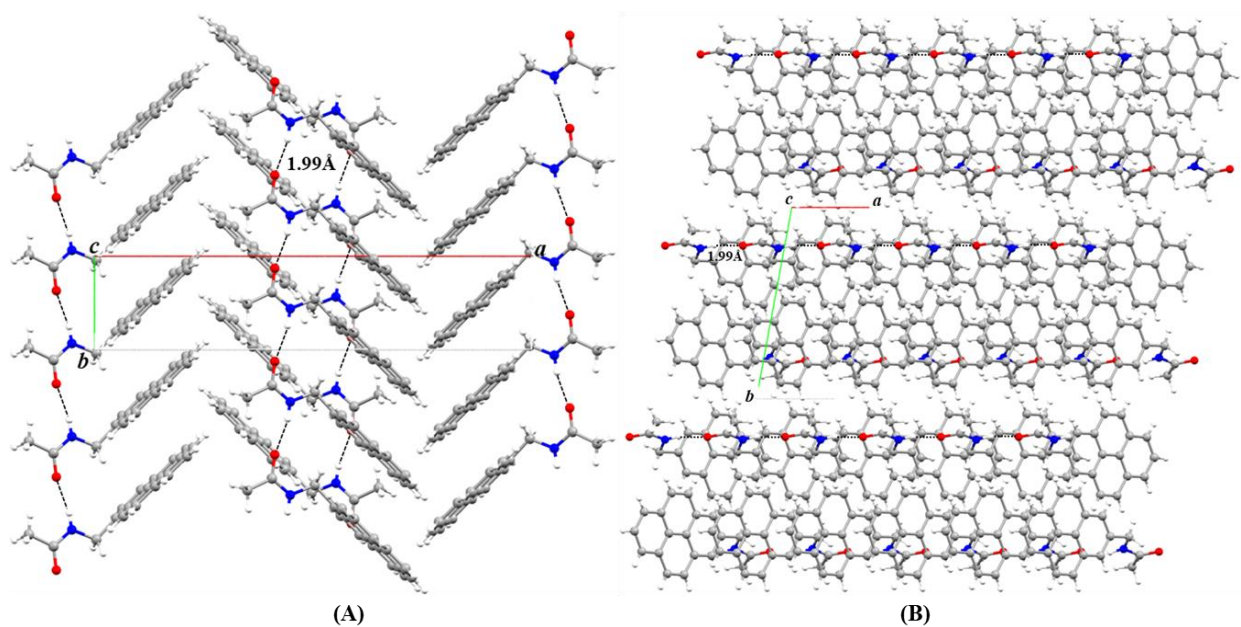

**Figure S9**. Packing of molecules down crystallographic '*c*' axis. A) **PyMA1**. B) **PyMA2** (H⋯O distance in N-H⋯O hydrogen bond is provided in both the cases).

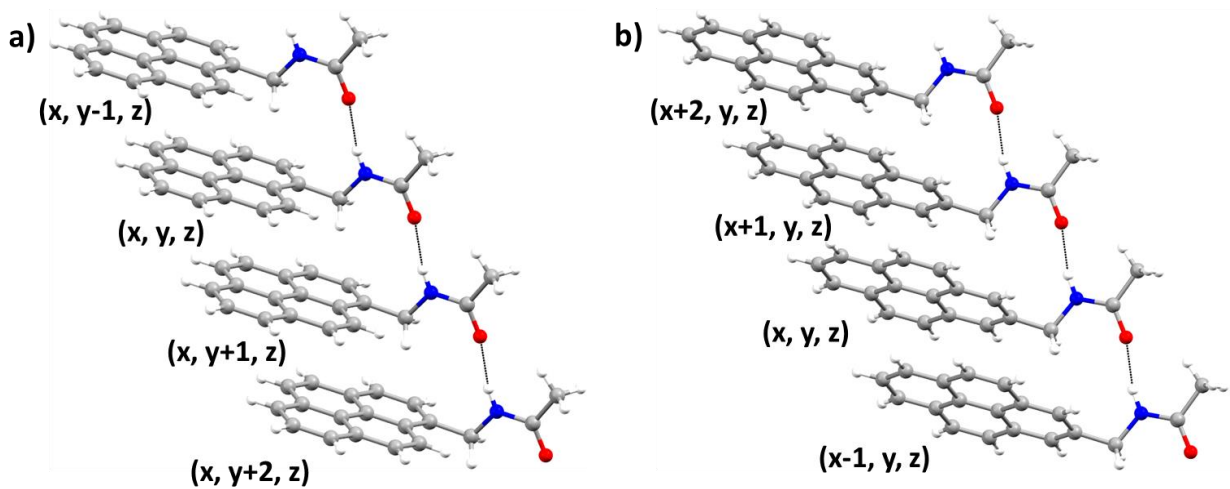

**Figure S10**. One dimensional π⋯π stacking between unit translated molecules empowered by strong intermolecular NH⋯O hydrogen bond along crystallographic axis '*b*' and '*a*' in PyMA1 and PyMA2, respectively.

**Table S6.** The lifetimes and the population percentages of the emissive species in solution and single crystals.

| Compounds | Concentration (mM) | **Solution** | | | |
| --- | --- | --- | --- | --- | --- |
| | | **(Monomeric fluorescence)** | | **(Excimer emission)** | |
| | | Population (%) | $\tau_1$ (*ns*) | Population (%) | $\tau_2$ (*ns*) |
| PyMA1 | 0.1 | **100** | (1.590 ± 0.003) | | |
| | 1.0 | **99.98** | (1.929 ± 0.002) | 0.02 | (29.017 ± 0.063) |
| PyMA2 | 0.1 | **100** | (0.921 ± 0.001) | | |
| | 1.0 | **100** | (1.210 ± 0.003) | | |
| **Single Crystals** | | | | | |
| Compounds | Temperature (K) | Population (%) | $\tau_1$ (*ns*) | Population (%) | $\tau_2$ (*ns*) |
| PyMA1 | 296 | 99 | (4.569 ± 0.005) | 1 | (35.591 ± 0.564) |
| | 230 | 96 | (4.805 ± 0.006) | 4 | (27.549 ± 0.166) |
| | 180 | 96 | (4.565 ± 0.006) | 4 | (26.826 ± 0.126) |
| | 130 | 96 | (4.048 ± 0.004) | 4 | (22.962 ± 0.076) |
| | 77 | 95 | (4.273 ± 0.005) | 5 | (23.512 ± 0.068) |
| PyMA2 | 296 | 94 | (7.306 ± 0.010) | 6 | (28.658 ± 0.167) |
| | 230 | 94 | (6.623 ± 0.009) | 6 | (30.341 ± 0.146) |
| | 180 | 93 | (6.064 ± 0.007) | 7 | (30.416 ± 0.110) |
| | 130 | 92 | (5.525 ± 0.007) | 8 | (29.466 ± 0.084) |
| | 77 | 89 | (5.210 ± 0.008) | 11 | (27.610 ± 0.070) |

\* Excitation wavelength=375nm

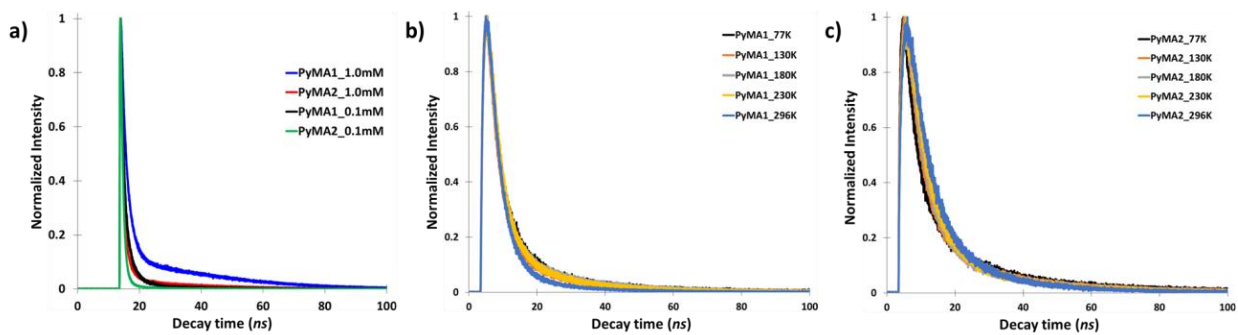

**Figure S11**. Emission lifetimes obtained from TCSPC measurements. a) PyMA1 and PyMA2 in Toluene. b) **PyMA1** and c) **PyMA2** in single crystals, at different temperatures.

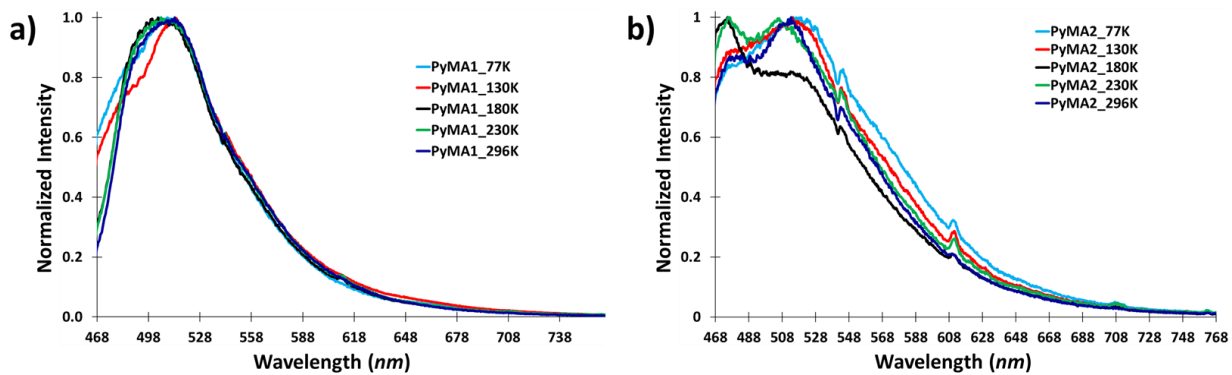

**Figure S12**. Emission spectra of a) PyMA1 and PyMA2 in single crystals at different temperatures.

**Table S7:** TDDFT-Calculated Main Orbital Contributions to Selective Singlet-Singlet Transitions in PyMA1 and PyMA2.

| Main orbital contributions | E (eV) | λ (nm) | f × 10$^{4}$ [a] |
|---|---|---|---|
| **PyMA1** | | | |
| H-1->L+1 (78%) | 5.36 | 231.34 | 6544 |
| H-5->L+2 (27%), H-4->L+3 (20%), H-3->L+3 (14%), H-2->L+5 (17%) | 7.00 | 177.04 | 6021 |
| H-4->L+2 (51%), H-3->L+2 (15%) | 6.42 | 193.09 | 5911 |
| **Excitation Transition** [b] | | | |
| HOMO->LUMO (92%) | 3.59 | 345.14 | 2982 |
| **PyMA2** | | | |
| H-1->L+1 (79%) | 5.20 | 238.44 | 9371 |
| H-4->L+2 (22%), H-3->L+2 (33%), H-1->L+4 (23%) | 6.48 | 191.44 | 6540 |
| H-5->L+2 (27%), H-4->L+5 (17%), H-3->L+3 (28%) | 7.07 | 175.27 | 5954 |
| **Excitation Transition** [b] | | | |
| HOMO->LUMO (89%) | 3.67 | 337.68 | 2674 |

[a] Oscillator strength.
[b] Main orbital contribution for the singlet-singlet HOMO-LUMO excitation transition.

**Table S8**: Decomposition of MOs into percentage contributions of pyrene and N-methylacetamide groups in Pyrene-1-methylacetamide **1** and Pyrene-2-methylacetamide **2**.

| Orbital type | Orbital energy (eV) | | Decomposition of molecular orbitals into element percentage contributions (%) | | | |
|---|---|---|---|---|---|---|
| | | | Pyrene | | N-methylacetamide | |
| | **1** | **2** | **1** | **2** | **1** | **2** |
| **LUMO+4** | 0.54 | 0.50 | 14 | 16 | 86 | 84 |
| **LUMO+3** | 0.14 | 0.19 | 92 | 99 | 8 | 1 |
| **LUMO+2** | -0.39 | -0.41 | 99 | 100 | 1 | 0 |
| **LUMO+1** | -0.97 | -1.00 | 99 | 96 | 1 | 4 |
| **LUMO** | -1.83 | -1.78 | 98 | 100 | 2 | 0 |
| **HOMO** | -5.58 | -5.61 | 97 | 100 | 3 | 0 |
| **HOMO-1** | -6.50 | -6.39 | 98 | 92 | 2 | 8 |
| **HOMO-2** | -7.00 | -6.93 | 19 | 2 | 81 | 98 |
| **HOMO-3** | -7.08 | -7.13 | 41 | 81 | 59 | 19 |
| **HOMO-4** | -7.16 | -7.22 | 59 | 33 | 41 | 67 |

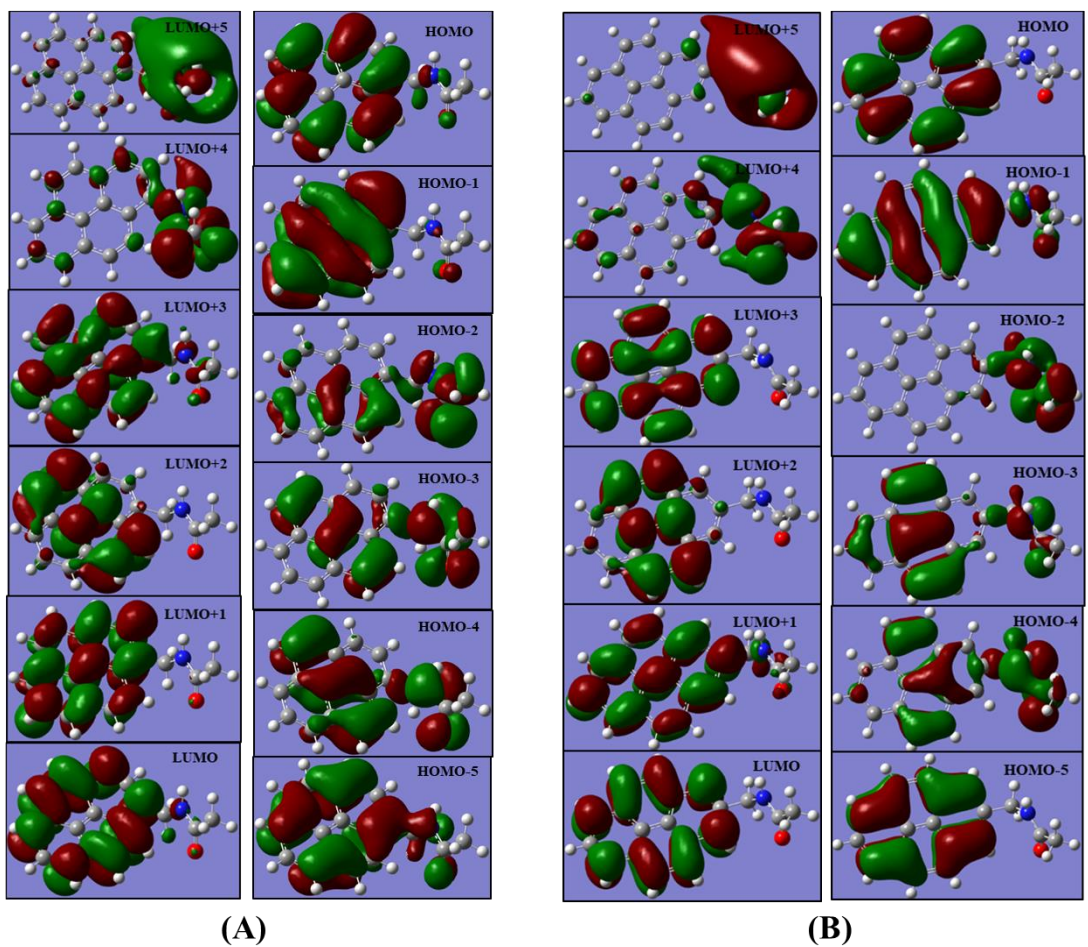

**Figure S13**. Selected frontier MOs involved in intense transitions. (A) **PyMA1**. (B) **PyMA2**.

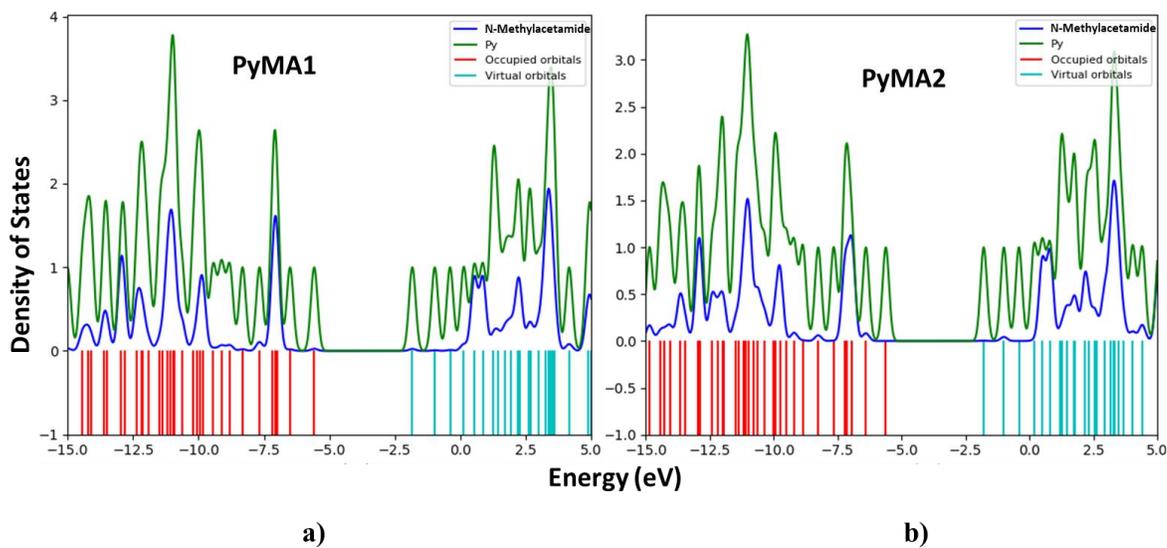

**Figure S14**. Density of States (DOS). a) PyMA1 and b) PyMA2. [Py: Pyrene]

**Table S9.** Decay time and population percentages of transient species observed in global analyses of *fs*-TA measurements.

| | *fs*-TA | | | | | | |
|---|---|---|---|---|---|---|---|
| | **PyMA1** | | | | | | |
| Conc. (mM) | Right singular vectors | $P_1$ (%) | $\tau_1$ (*fs*) (IC) | $P_2$ (%) | $\tau_2$ (*ps*), probably $S_0 \leftarrow S_2$ fluorescence | $P_3$ (%) | $\tau_3$ (*ps*), probably $T_2 \leftarrow S_2$ ISC |
| 1.0 | 595nm (V1) | **99.3** | **(163 ± 7)** | **0.4** | **(1050 ± 190)** | **0.3** | **(31.4 ± 3.6)** |
| | 520nm (V2) | Decayed partially | | | | | |
| | (V3)[a] | Decayed partially | | | | | |
| | **PyMA2** | | | | | | |
| 1.0 | 580nm (V1) | **93.6** | **(754 ± 54)** | **3.4** | **(519.76 ± 145.91)** | **3.0** | **(45.4 ± 16.4)** |
| | 530nm (V2)[b] | Decayed partially | | | | | |
| | 485nm (V3) | Decayed partially | | | | | |

[a] The vector V3 has three broad bands centered on 520nm, 600nm and 660nm. [b] The vector V2 has two prominent shoulders at 500nm and 605nm.

**Table S10.** Decay time and population percentages of principal components obtained from SVD analyses of *ns*-TA measurements.

| | *ns*-TA | | | | |
|---|---|---|---|---|---|
| | **PyMA1** | | | | |
| Conc. (mM) | Right singular vectors | $P_1$ (%) | $\tau_1$ (*ns*) | $P_2$ (%) | $\tau_2$ (*ns*) |
| 1.0 | 480nm (V1) | **75** | **21.7 ± 0.2** | **25** | **254.8 ± 3.9** |
| | 425nm (V2) | | **238.8 ± 0.9** | | |
| | 525nm (V2) | **72** | **240.3 ± 1.8** | **28** | **28.3 ± 0.7** |
| 0.1[a] | 480nm (V1) | **71** | **18.5 ± 1.1** | **29** | **235.4 ± 19.1** |
| | 425nm (V2) | | **251.6 ± 4.7** | | |
| | 525nm (V2) | **69** | **243.5 ± 8.4** | **31** | **18.3 ± 2.0** |
| | **PyMA2** | | | | |
| 1.0 | 480nm (V1) | **78** | **20.2 ± 0.2** | **22** | **262.9 ± 5.7** |
| | 425nm (V2) | | **258.2 ± 1.2** | | |
| | 530nm (V2) | **72** | **252.0 ± 1.8** | **28** | **21.0 ± 0.6** |
| 0.1[a] | 485nm (V1) | **82** | **15.4 ± 0.8** | **18** | **239.7 ± 36.1** |
| | 425nm (V2) | | **268.3 ± 8.3** | | |
| | 528nm (V2) | **61** | **259.2 ± 10.3** | **39** | **6.4 ± 0.9** |

[a] The TA data sets correspond to 0.1mM concentration for PyMA1 and PyMA2 are not presented here. Only the decay lifetime for several transient species represented by associated vectors obtained from SVD analysis is presented.

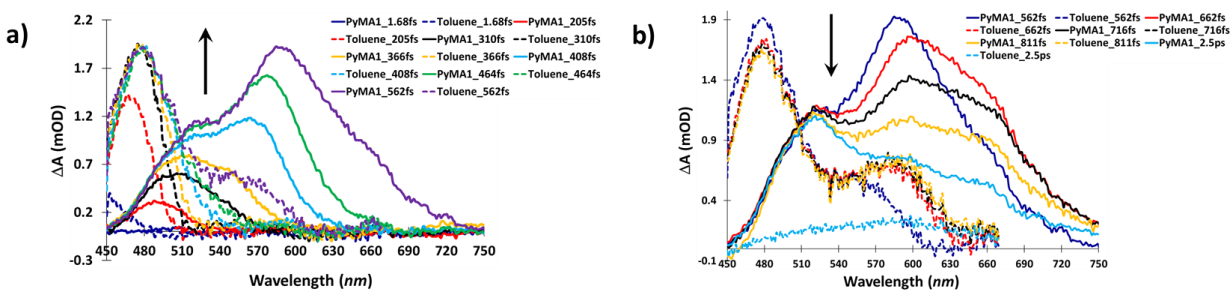

**Figure S15**. *fs*-TA of toluene (dashed lines) and PyMA1 (solid lines), a) up to 562*fs*. b) From 562*fs* to 2.5*ps*.

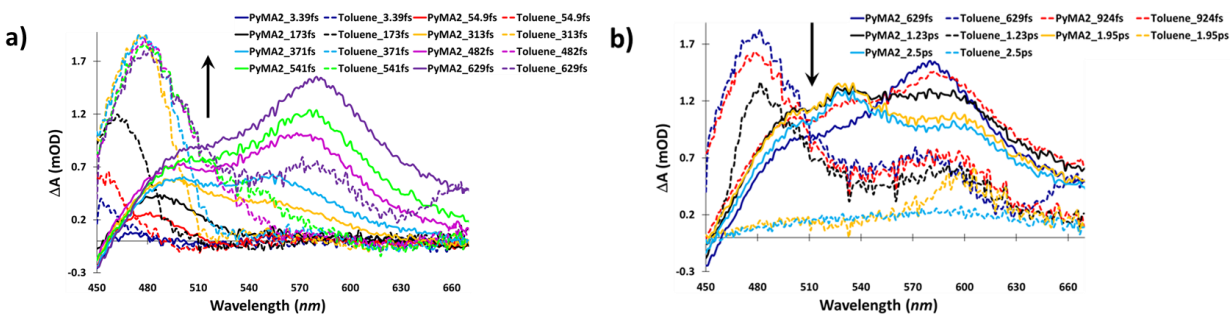

**Figure S16**. *fs*-TA of toluene (dashed lines) and PyMA2 (solid lines), a) up to 629*fs*. b) From 629*fs* to 2.5*ps*.

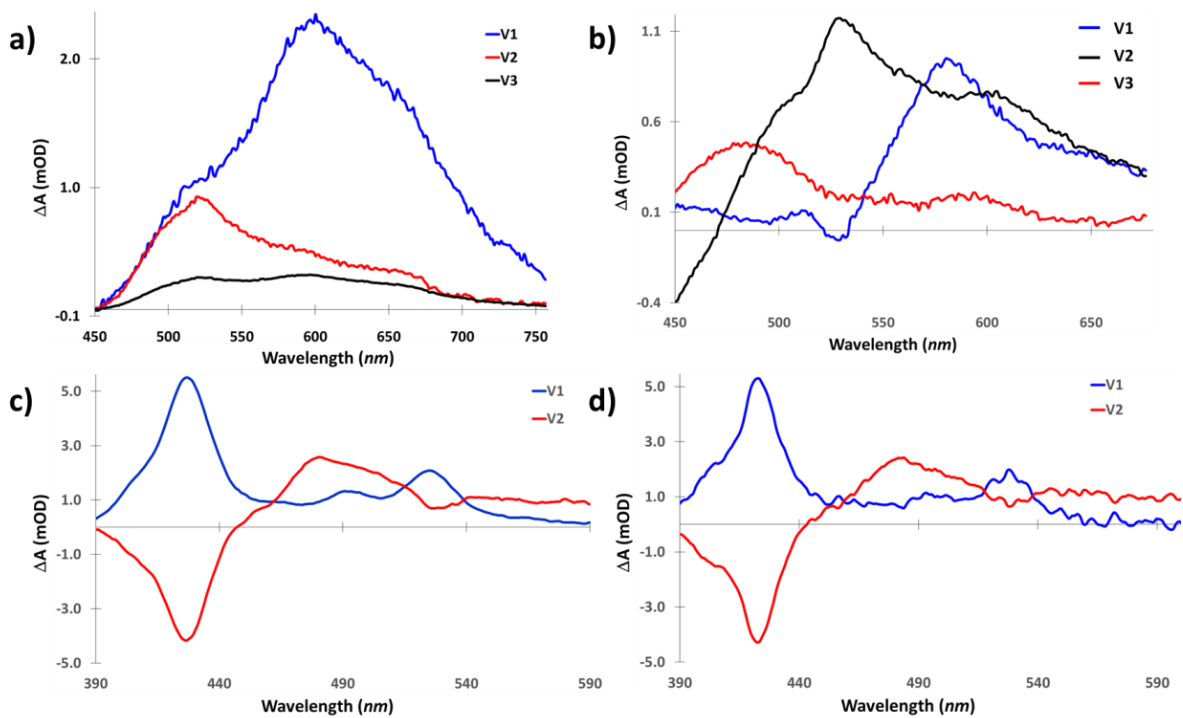

**Figure S17**. Right singular vectors obtained from global analysis for PyMA1 and PyMA2 at 1.0mM, from *fs*-TA and *ns*-TA datasets. The *fs*-TA of a) PyMA1 and b) PyMA2. The *ns*-TA of c) PyMA1 and d) PyMA2.

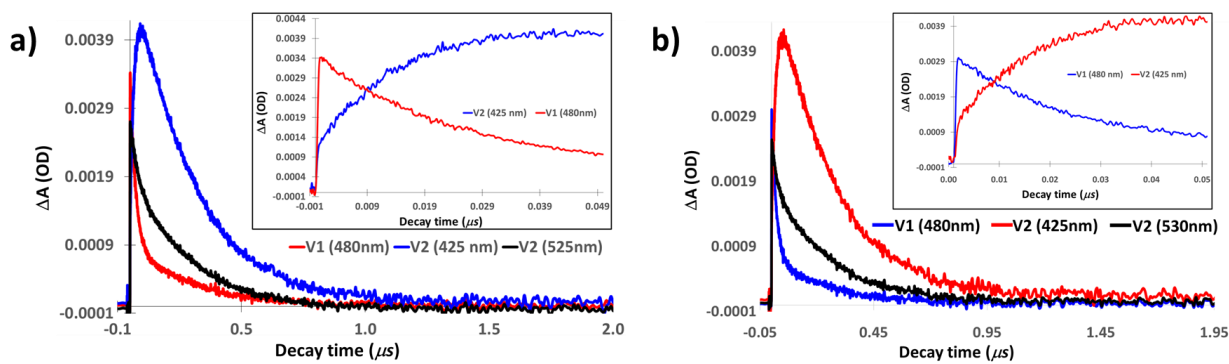

**Figure S18**. Kinetic profiles of a) PyMA1 and b) PyMA2, at 1.0mM, for principal components obtained from SVD analysis of *ns*-TA datasets.